\documentclass[11pt]{article}
\usepackage{graphicx}
\usepackage{url}
\usepackage{makecell}
\usepackage{xcolor}
\usepackage{framed}
\usepackage{caption}
\usepackage{subcaption}
\usepackage{float}
\usepackage{tcolorbox}
\usepackage{textcomp}
\usepackage{hyperref}
\usepackage{multicol}
\usepackage{blindtext}
\usepackage{amssymb}
\usepackage{booktabs} 
\usepackage[colorinlistoftodos]{todonotes}
\usepackage{geometry}
 \let\OLDthebibliography\thebibliography
\renewcommand\thebibliography[1]{
  \OLDthebibliography{#1}
  \setlength{\parskip}{0pt}
  \setlength{\itemsep}{0pt plus 0.5ex}
}
 
\definecolor{lightgray}{gray}{0.94}

\newcounter{mainfindingid}
\newcommand{\mainfinding}[2]{\refstepcounter{mainfindingid}\label{#1}\item[
	\ifthenelse{\value{mainfindingid}=1}{Observation-}{O-}\arabic{mainfindingid}:
	] #2}

\usepackage{hyperref}
\hypersetup{
    colorlinks=true,
    linkcolor=black,
    filecolor=red,      
    urlcolor=red,
    citecolor=blue
}

\usepackage[format=plain,
            labelfont={bf,it},
            textfont={bf,it}]{caption}

\begin{document}

\thispagestyle{empty}

\begin{center}

Vrije Universiteit Amsterdam

\vspace{1mm}

\includegraphics[height=28mm]{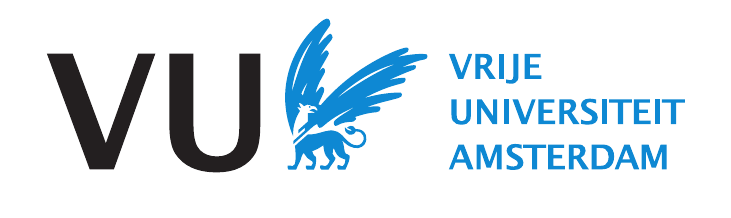}

\vspace{1.5cm}

{\Large Bachelor Thesis, Project Report}

\vspace*{1.5cm}

\rule{.9\linewidth}{.6pt}\\[0.4cm]
{\huge \bfseries Characterizing User and Provider Reported Cloud Failures\par}
\rule{.9\linewidth}{.6pt}\\[1.5cm]

\vspace*{2mm}

{\Large
\begin{tabular}{l}
{\bf Author:} ~~Mehmet Berk Cetin~~~~ (2644886)
\end{tabular}
}

\vspace*{1.5cm}

\begin{tabular}{ll}
{\it 1st supervisor:}   & ~~prof. dr. ir. Alexandru Iosup \\
{\it 2nd supervisor:}   & ~~dr. ir. Animesh Trivedi \\ 
{\it daily supervisor:} & ~~ir. Sacheendra Talluri \\

\end{tabular}

\vspace*{1cm}

\today\\[4cm] 
\end{center}
\newpage

\renewcommand*\contentsname{Table of Contents\newline}
\tableofcontents
\newpage

\begin{abstract}
Cloud computing is the backbone of the digital society. Digital banking, media, communication, gaming, and many others depend on cloud services. Unfortunately, cloud services may fail, leading to damaged services, unhappy users, and perhaps millions of dollars lost for companies. Understanding a cloud service failure requires a detailed report on why and how the service failed. Previous work studies how cloud services fail using logs published by cloud operators.
However, information is lacking on how users perceive and experience cloud failures. Therefore, we collect and characterize the data for user-reported cloud failures from Down Detector for three cloud service providers over three years. We count and analyze time patterns in the user reports, and derive failures from those user reports and characterize their duration and interarrival time.
We characterize provider-reported cloud failures and compare the results with the characterization of user-reported failures. The comparison reveals the information of how users perceive failures and how much of the failures are reported by cloud service providers. 
Overall, this work provides a characterization of user- and provider-reported cloud failures and compares them with each other.
\end{abstract}

\section{Introduction} \label{sec:introduction}
Cloud computing systems are the new computing paradigm in the 21st century~\cite{buyya2009cloud}. These computing systems enable consumers to satisfy their excessive computing needs that could not be fulfilled by personal computers~\cite{maguire_vendors}.
Cloud computing services are important for digital banking, communication (e.g., WhatsApp), entertainment (e.g., online gaming), multimedia (e.g., Netflix) scientific research (e.g., Google Colab), online education (e.g. Zoom) and other purposes. 
The digital society depends on the cloud systems and the services they offer. Unfortunately, cloud services fail from time to time, causing the services that depend on the cloud service to also fail. These failures can be mild, such as a single-region outage affecting only the users at that region, or severe, such as a multi-region outage of multiple services, affecting much more users in multiple locations. 
Moreover, cloud failures hurt the customers, and they cause financial and reputation damages to the cloud service providers. Cloud service downtime for a couple of minutes can cause millions of dollars loss in revenue~\cite{googleoutage2013, facebookoutage2014, blackberryoutage2014, aws_outage}.

Information about cloud service failures is important. Companies like Amazon, Microsoft, and Google offer status updates for many of their cloud services.
To understand and prevent future cloud failures, we need cloud failure reports, which are used to identify how, when, and why the service failure occurred. Unfortunately, According to preliminary data from users, cloud vendors do not report all failures.~\cite{hacker_news}. 
Nonetheless, crowdsourcing failure aggregators can capture failure events that were not reported by the vendors using user failure reports. We can gain invaluable information on cloud service failures by investigating the failure reports offered by the failure aggregators.
Therefore, unlike previous work, which focuses on failure reports that were self-reported or logged by cloud operators~\cite{el2017learning, javadi2013failure, birke2014failure}, this work focuses on analyzing cloud failures reported by users and compares them with cloud failures reported by the providers. 

In this work, we investigate the failures of back-end cloud services offered by Amazon Web Services~(AWS), Google Cloud Platform~(GCP), and Microsoft Azure.
We collect the data of user-reported cloud failures for AWS, GCP, and Azure from a crowd-sourced failure aggregator, namely Down Detector. 
Afterwards, we characterize the user-reported data using basic statistical methods. 
We count and analyze weekly and monthly time patterns in these user reports, derive failure events from them, and characterize their duration and interarrival time.
We also characterize previously scraped data of provider-reported cloud failures and compare the analysis of user- and provider-reported cloud failures with each other.

There are two main contributions in this paper.
We analyze data of user- and provider-reported cloud failures and compare the failure reports with each other for the biggest cloud vendors in the market. 
We open-source data on user-reported cloud failures from two separate sources, mainly Down Detector and Outage Report~\cite{outagereport}, enabling other researchers to run our scripts and use the data for further research.

\section{Background Information} \label{sec:background}

\subsection*{Cloud Services} \label{subsec:cloud-services}
A cloud service provides access to a data center infrastructure that a host maintains. Cloud services can be used as Infrastructure as a Service (scalable computing resources), Software as a Service (cloud-based software services), or Platform as a Service (cloud environment to develop, manage, and host applications). The providers are responsible for all management, maintenance, security, and upgrades for the cloud services~\cite{cloud_services_background}. 

Cloud services can depend on other cloud services. For instance, Netflix, Airbnb, Disney, Reddit, Epic Games, and many more use Amazon Web Services~(AWS) to host their applications~\cite{awsusers, hunt_netflix_aws}. Due to this hierarchy, a cloud service failure in AWS can lead to many more service failures for the services that depend on AWS~\cite{aws_outage}. Therefore, a service failure in one major cloud service provider can lead to other service failures~\cite{microsoft_refund, microsoft_xbox, gcp_auth_failure}.

There are two types of cloud services: client-facing and back-end. 
Client-facing cloud services can be Netflix, YouTube, Instagram and many more applications. Back-end services can be Google Compute Engine, Amazon Elastic Computing Service, Microsoft Azure Storage, and many more services.
The top three major cloud providers are Amazon Web Services (AWS), Google Cloud Platform (GCP), and Microsoft Azure~\cite{major_cloud_services}. 

\subsection*{User-Reported Cloud Failures} \label{subsec:user-cloud-failures}
We collect the data of user-reported cloud failures from Down Detector.
Down Detector is a crowd-sourced failure aggregator which collects reports in 15 minute granularity. Users are the main sources for failure identification in Down Detector, nonetheless, there are other series of sources, such as Twitter~\cite{downdetector}.
Moreover, users who experience issues in any kind of cloud service can report the symptoms they're experiencing to Down Detector. When the number of reports show a rapid increase relative to the baseline, a failure event is detected and briefly published on a separate page for the cloud service that is experiencing a failure. From those failure pages, we extract single or multiple failure events because each vendor can contain more than one failure event.

\begin{figure}[H]
    \centering
    \includegraphics[width=\textwidth]{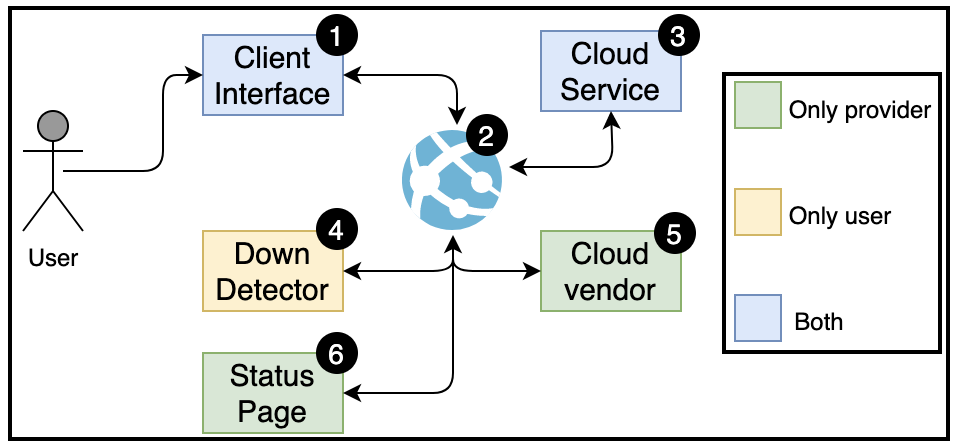}
    \caption{Failure model for user- and provider-reported cloud failures.}
    \label{fig:failure_model}
\end{figure}

In this work, the failure model for user-reported cloud failures is depicted in Figure~\ref{fig:failure_model}. The yellow box belongs only to the user-reported failure model, the green boxes belong only to the provider-reported failure model, and the blue boxes belong to both failure models.

The user-reported failure model is as follows: 
users access the cloud service (3) via an internet connection (2) using a client interface (1), which in AWS and Azure is a command line interface and in GCP is a programmatic interface. If a cloud service fails, users can experience symptoms and report them to crowdsourcing failure aggregators, which is Down Detector (4) in our case, using a browser connected to the internet. 

\subsection*{Provider-Reported Cloud Failures} \label{subsec:provider-cloud-failures}
We consider the failure model for provider-reported cloud failures to showcase the difference of how user- and provider-reported cloud failures are generated. The provider-reported failure reports we analyze in this work are scraped from the status pages~\cite{gcp_status_page, azure_status_page, aws_status_page} of the cloud vendors.

In Figure~\ref{fig:failure_model}, the failure model for provider-reported cloud failures is shown. Similar to user-reported cloud failure models, users access the cloud service (3) via an internet connection (2) using a client interface (1). If a cloud service fails, the cloud vendor (5) composes a report about the failure event and posts it in their status page (6). Users can access those reports via a browser connected to the internet.

\subsection*{Terms \& Definitions} \label{subsec:terms}
A cloud service \textit{failure} or \textit{outage} is a certain amount of time when the service isn't performing as expected or is completely unavailable. A cloud service failure is logged by a \textit{failure or outage report}, which is an implication by the user to the service provider about a cloud service failure. Failure \textit{duration} is the time difference between the start and end of a failure event. For the extraction method of the start and end times of a failure event, see Section~\ref{subsec:failure_extraction}. The \textit{interarrival time} is the amount of elapsed between two consecutive failure start times. We define an \textit{unavailable} cloud service as a service that is inaccessible by users until the failure is troubleshooted. During a failure, a cloud service might still be partially available, only experiencing \textit{performance degradation} because of various fault-tolerance methods. Many services depend on each other, and when the service that is used by many other services fails, multiple failures will happen, leading to \textit{multi-service failure}. 
\section{Problem Statement} \label{sec:problem}

Cloud services are an essential part of various applications~\cite{kiryakova2015application}. However, cloud services fail~\cite{gunawi2016does}, and these failures are not always reported by the company that provides these services~\cite{hacker_news}. Crowd-sourcing failure aggregators, which collect user failure reports, can help identify unreported failures and provide us with information on the cloud service failures.
Moreover, information is lacking on how users perceive and experience cloud failures. 
We cannot understand how and when cloud failures happen in the perspective of the users if we do not have information on them. 
To address this issue, we collect the data of user-reported failures from a crowdsourcing failure aggregator, namely Down Detector~\cite{downdetector}. We characterize user-reported failures and compare them with the characterization of a reliable source, provider-reported cloud failures.
After the characterization and comparisons, we try to answer the following question: \textit{How do cloud services fail from the perspective of the users?} From this main question, we derive three research questions: \\ \\
\textbf{RQ1:}~How to collect the data of user-reported cloud failures from Down Detector and Outage Report? \\
\textbf{RQ2:}~Are there any significant differences between user- and provider-reported cloud failures? \\
\textbf{RQ3:}~How can we characterize user- and provider-reported cloud failures? \\

\section{Methodology} \label{sec:methodology}

\subsection{Data Collection}
In this section, we address RQ1 and explain how we collected user-reported failure data and extracted failure events from the user reports. 
We also explain our failure analysis method used to process and create graphs for understanding the user- and provider-reported data.

\begin{figure}[H]
    \centering
    \includegraphics[width=\textwidth]{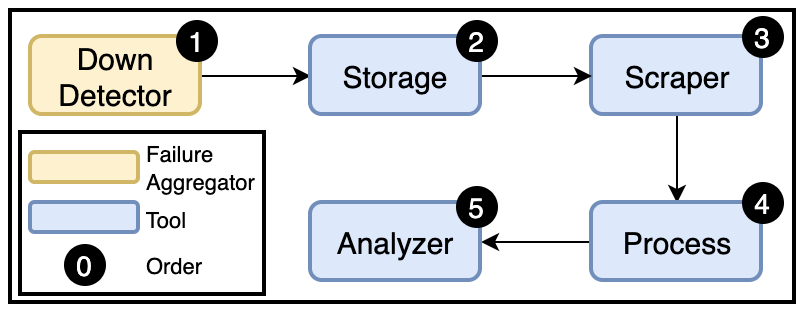}
    \caption{Data collection, processing, and analysis.}
    \label{fig:failure_process}
\end{figure}

We collected (scraped) the data of user-reported cloud failures for AWS, GCP, and Azure from Down Detector over a three-year period.
Figure~\ref{fig:failure_process} depicts the process of data collection, processing, and analysis.
To save the website from being overworked, we download each web page that contains the failure events from Down Detector (1) once a minute and store them locally (2). Afterwards, we locally extract the relevant data (report count and most reported problems) from the events we downloaded using various tools from Python (3). 
To save the extracted data in a processable format (4), we parse the data and save them as data frames.
The failure events are summarized in Table~\ref{tab:user-detailed-summary}. 
As for provider-reported cloud failures, the failure data for AWS, GCP, and Azure were already scraped from the status pages of the cloud vendors. We only processed (4) the data and analyzed (5) it.

\subsection{Failure Extraction} \label{subsec:failure_extraction}
Down Detector identifies failure events according to the increase in the number of user reports. The failures are presented per page, see Section~\ref{subsec:user-cloud-failures}. In each page, the number of user reports are monitored for 23 hours and 45 minutes. Inside this time frame, Down Detector sets a failure start time using an algorithm. 
Nonetheless, there may be multiple failures per page.
Therefore, we use a method, which we got inspired from the alternative identification swarm-size-based algorithm of Zhang et al.~\cite{zhang2011identifying}, to extract multiple failure events per page. 
We set a threshold per page and extract the start time of a failure event when the number of reports exceed the threshold and extract the end time when the number of reports drop below the threshold. 
Per page, we determine the 75th quantile value and set that as the threshold for that page.
The failure events were already distinct in the data of provider-reported cloud failures.

\subsection{Failure Analysis Method}
After we collect and process the data of cloud failures (see Figure~\ref{fig:failure_process}), we analyze the processed data (5) using basic statistical methods like averaging and aggregation. We analyze user-reported failures by weekly, monthly, yearly, and seasonal granularity. For the weekly analysis, we got inspired from the analysis conducted in the Workflow Trace Archive~\cite{versluis2020workflow}. Moreover, we imitate some methods and figures in another research very similar to ours ~\cite{sacheen_user_reports}.
We compare how many of the user- and provider-reported failures overlap with the goal of observing how failures self-reported by AWS, Azure, and GCP differentiate from user-reported failures. 
We compare the failure event start time, event count, duration, and interarrival times to identify the differences of the nature of the failures for user- and provider-reported cloud failures. For instance, using the graph that depicts failure duration, we observe that user-reported failure events last longer than provider-reported failure events.

\section{Analysis of User-Reported Cloud Failures} \label{sec:user_analysis}


In this section, we present our findings by analyzing user-reported cloud service failures and address \textit{RQ3}.
We use the data we collect from Down Detector on Amazon Web Services (AWS), Microsoft Azure and Google Cloud Platform (GCP). 
The analysis covers the failure user report count, event count, duration, and interarrival time.
After the analysis of user-reported cloud failures, we analyze, in Section~\ref{sec:provider_analysis}, provider-reported cloud failures for the same vendors over the same period. This gives us the opportunity to compare user- and provider-reported cloud failures and see how they match or differentiate. 
The characterization of both user- and provider-reported failures cover the event count, duration, and interarrival time of the failures. 
At the beginning of each subsection we present our main observations.


\begin{table}[ht]
\centering

\setlength{\fboxsep}{0pt}
\resizebox{\textwidth}{!}{
\colorbox{lightgray}{
\begin{tabular}{lrrrr}
\toprule
              Cloud services &  \# failure reports &  \# failure events & Date range\\
\midrule
   Google Cloud Platform &        10440 &           42 &    12 Nov. 2018 - 14 Dec. 2020\\
   Microsoft Azure &         92856 &           314 &    4 Jan. 2018 - 17 Dec. 2020 \\
   Amazon Web Services &         118415 &           452 &     4 Jan. 2018 - 19 Dec. 2020\\
 
\midrule
Total & 221711 & 808 \\
\bottomrule

\end{tabular}}
}

\caption{\label{tab:user-detailed-summary}Detailed summary for User-Reported Cloud Failures, separated by cloud services.}
\end{table}

There are in total 221711 user failure reports and 808 failure events in the data we collected. The number of reports for all the cloud services is summarized in Table~\ref{tab:user-detailed-summary}. The \textit{\# failure reports} contains the count for the total number of user failure reports across the whole data set. The \textit{\# failure events\"} contains the count for the total number of failure events extracted using the method described in Section~\ref{sec:methodology}. The \textit{Date range} contains the date for the first and last reports recorded for each cloud vendor.

\subsection{User Report Counts} \label{subsec:user:failure_report_trends}
In this section, we analyze the weekly, monthly, and seasonal user report counts.  
The analysis of failure reports reveals information on the nature of the failure event, such as when failures start and whether peaks overlap with failure events proclaimed by cloud vendors. We intend to see the general distribution of user reports and determine in what part of the week, month, and year failures happen in the perspective of users. Moreover, we use the reports to create graphs, depicting certain peaks that match with failures affirmed by cloud vendors. 
Our main findings are below.

\begin{tcolorbox}
\begin{description}
\itemsep-0.2em
	\mainfinding{analysis:mf:user:weekly_peak}{All peaks in the weekly average report counts occur during the evening of the day.}
	\mainfinding{analysis:mf:peak_match}{The peaks for cloud vendors in the average monthly report count graph match with failures proclaimed by cloud vendors.}

\end{description}
\end{tcolorbox}

In Table~\ref{tab:user-detailed-summary}, there are differences in report counts between cloud vendors. The report count of GCP is 9 times less than Azure and about 11 times less than AWS. AWS has the highest report count with 118415, and is approximately 26000 reports higher than Azure. This implies that either the users of AWS use Down Detector to report failure events more than the Azure's and GCP's users, or simply AWS failed more than Azure and GCP and again lead to more user reports in Down Detector. 
In 2018, according to estimates~\cite{cloud_market_shares}, more than half of the total cloud market share is owned by AWS, Azure and GCP. AWS owns around 33\% of the market share while Azure owns 15\%. GCP occupies only 5\% of the market share. Moreover, during the first quarter of 2021, it is estimated~\cite{cloud_market_shares_21} that the market share for GCP has increased by 2\%, resulting in a market share of 7\%. Azure has grown to 19\% while AWS has relatively decreased down to 32\%. These cloud infrastructure market shares imply that AWS is one of the largest cloud providers in the planet and has more users than Azure and GCP, leading to more reports when an outage happens. 

\begin{figure}[H]
    \centering
    \includegraphics[width=\textwidth]{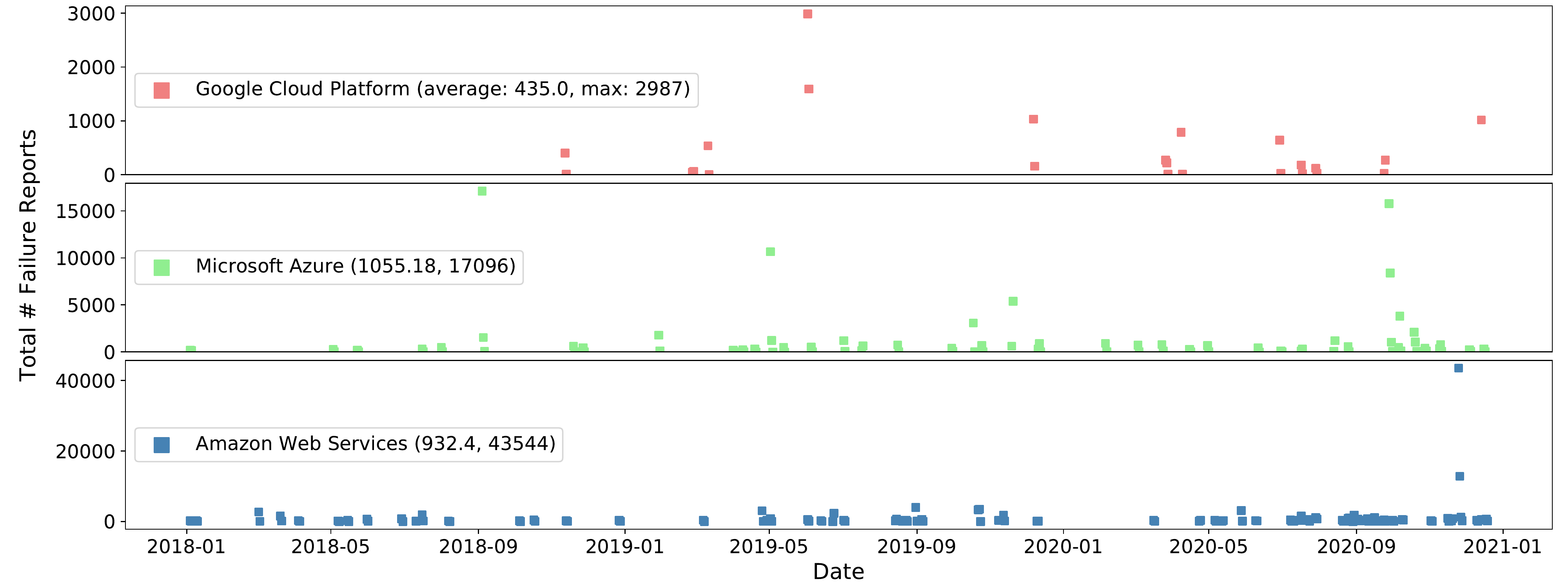}
    \caption{Number of failure reports aggregated per day, separated by cloud vendors.}
    \label{fig:user_general_report_count_analysis}
\end{figure}

We first investigate the distribution of user reports over the whole season. Our main intention is to observe the general overview of the failure report distribution over the whole season. We achieve that by creating the seasonal failure report count graph shown in Figure~\ref{fig:user_general_report_count_analysis}. 
The vertical axis depicts the total number of failures per day. The horizontal axis depicts the date in the granularity of one day.
According to the figure, Azure has the highest average report count, and AWS has the highest maximum report count in a day. 
Failure events with low user report counts might have decreased the average report count for AWS. Nonetheless, the highest total number of user reports belongs to AWS, indicating that a couple of extreme outliers have increased the total report count for AWS.
Azure has, in average, a higher average report count but lower total report count than AWS.  
In contrast to AWS and Azure, the user reports for GCP start at the end of 2018. When compared to the other two cloud services, around 11 months of failure reports are missing for GCP. Therefore, the average and maximum report counts for GCP are lower than AWS and Azure. 

\begin{figure}[H]
    \centering
    \includegraphics[width=\textwidth]{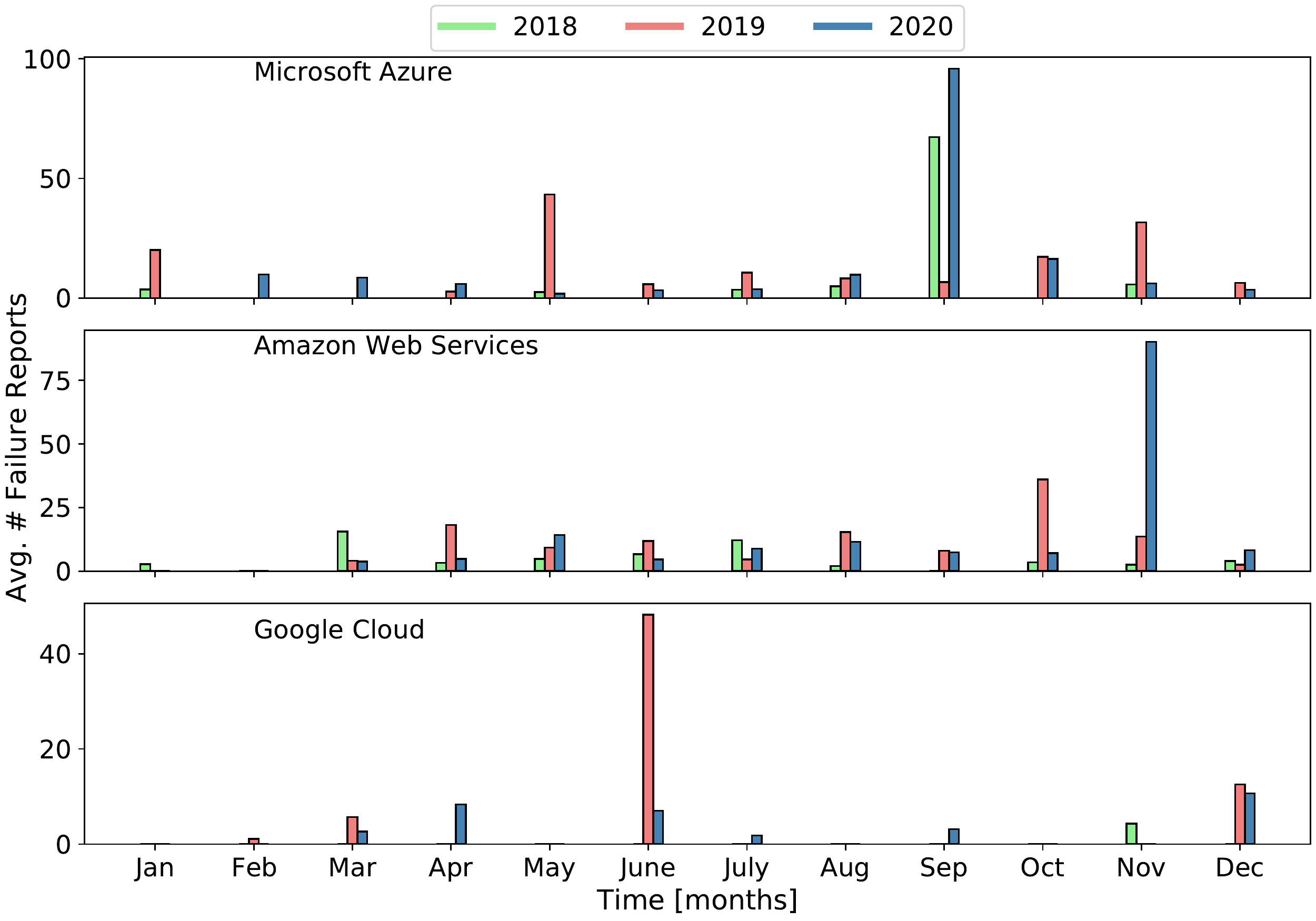}
    \caption{Average report count per month over three year, separated by cloud vendors.}
    \label{fig:user_monthly_yearly_average_report_count}
\end{figure}

We investigate the average user report count per month over three years.
Our goal is to see if there are any cloud vendor reports during the month that have high failure count. 
Therefore, we create a graph depicting the average number of the monthly user report counts over three years shown in Figure~\ref{fig:user_monthly_yearly_average_report_count}.
The vertical axis depicts the number of reports averaged per month on a yearly basis. The horizontal axis depicts the months in a year. According to the figure, within GCP, the highest average report count is in June 2019. The official report from Google~\cite{gcp_auth_failure} indicates that there was a disruption in Google’s network, which caused slow performance and elevated error rates on several Google services, including GCP. A 30\% reduction in network traffic was experienced in GCP.

The highest average report count for AWS is in November 2020. According to an official AWS report~\cite{aws_nov_2020_outage}, there was a failure event that lasted for approximately an hour. The outage effected Internet of Things (IoT) devices and online services. The outage was caused by a disruption in Kinesis, which is tasked with the job of collecting and analyzing real-time streaming data on AWS. Small capacity, which AWS does not reveal the amount, was added to Kinesis. The new capacity, however, has caused all the servers in the fleet to exceed the maximum number of threads allowed by an operating system configuration, which was the root cause for the disruption in Kinesis. Furthermore, the second highest average report count within AWS occurred in October 2019. There is no official report from the AWS status page indicating an outage. Other sources~\cite{aws_oct_2019_ddos, aws_oct_2019_ddos2, aws_oct_2019_ddos3} reported that the infrastructure of AWS was hit by a DDoS attack. This lead to the failures of many other services and applications that rely on AWS.

The highest average report count for Azure is in September 2020. At the end of that month, according to official reports~\cite{azure_sep_2020_auth}, Azure experienced an outage that lasted for approximately 3 hours. Customers of Microsoft encountered authentication errors across multiple Microsoft services and Azure Active Directory~(Azure AD) integrated applications, leading to users not being able sign into Microsoft and third-party applications, which use Azure AD for authentication. 

\begin{figure}[H]
    \centering
    \includegraphics[width=\textwidth]{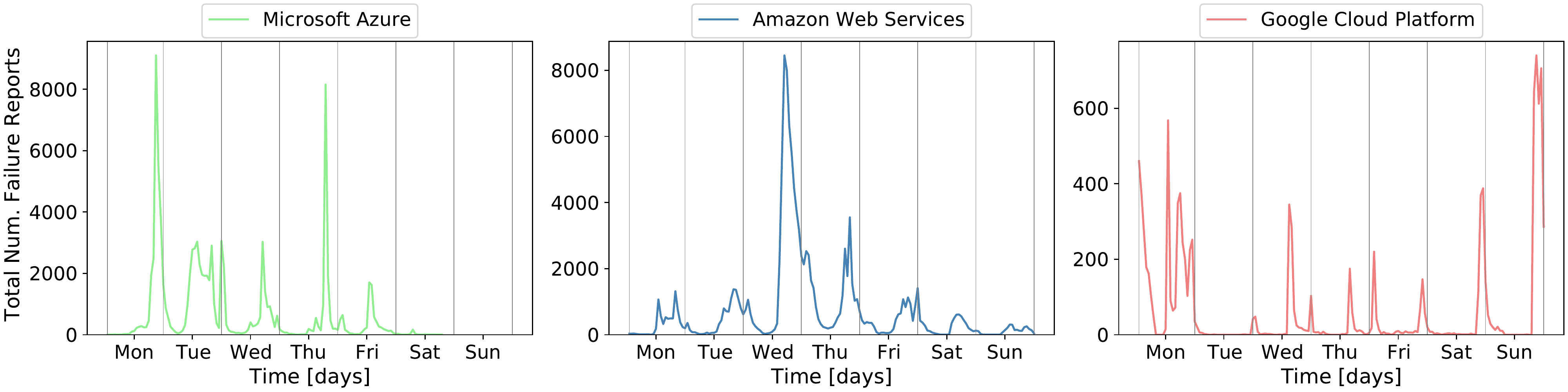}
    \caption{Distribution of failure reports aggregated by hour of week, separated by cloud vendors.}
    \label{fig:weekly_total_report_count}
\end{figure}

We investigate failure reports for the hours of the week. Our intention is to find out whether users are more likely to report failures during certain parts of a day and week than others. To understand the weekly user report trends, we create a graph that shows the weekly number of reports for each cloud service in Figure~\ref{fig:weekly_average_report_count}. 
The total number of failure reports in the vertical axis are calculated in the following format. We group the data points by day of the week and hour of day. Afterwards, we take the sum for each data point within its corresponding group. There are 168 data points, since each week is 7 days and each day is 24 hours, resulting in 168 hours per week. The horizontal axis depicts the days of the week and the times are arranged according to the Greenwich Mean Time Zone (Coordinated Universal Time). 

In the figure, on weekends, users didn't report failures for AWS and Azure. This implies that either AWS and Azure did not fail much on the weekend, leading to no failure symptoms and almost zero user reports, or the services AWS and Azure offer are mostly work related, and the users did not use the services because they mostly do not work on the weekend, and did not experience any failure symptoms and report them to Down Detector.
Moreover, all global report count peaks occur during the evening of the day (O-\ref{analysis:mf:user:weekly_peak}). The peak during Monday for Azure matches major outages that also occur during the evening of that day~\cite{azure_sep_2020_auth} (see 19 October 2020 and 28 September 2020). The peak during Wednesday for AWS matches a major outage that started around noon in the UTC timezone and continued to effect AWS during the evening~\cite{aws_nov_2020_outage}.
On Sunday evening, which there is peak at that time for GCP, a disruption in Google’s network caused performance degradation for GCP~\cite{gcp_june_2019_outage}. 

\subsection{Failure Event Counts} \label{subsec:user:failure_count}
In this section, we observe the number of user-reported cloud failure events per month and year. 
We explain how we extract the failure events in Section~\ref{subsec:failure_extraction}. 
Our goal is to observe the differences of the number of failure events between cloud vendors. We also use the results of user-reported failures and compare them with provider-reported cloud failures to see the similarities and differences, see Section~\ref{subsec:provider:failure_events}. 
Our main observations for this section are below.

\begin{tcolorbox}
\begin{description}
\itemsep-0.2em
    \mainfinding{user:failure_event:mf:gcp_less}{The number of failure events for AWS and Azure are more than GCP.}
    \mainfinding{user:failure_event:mf:extreme_user_report}{Certain failure events have extreme number of user reports, causing peaks in Figure~\ref{fig:user_monthly_yearly_average_report_count}}
\end{description}
\end{tcolorbox}

\begin{figure}[H]
    \centering
    \includegraphics[width=\textwidth]{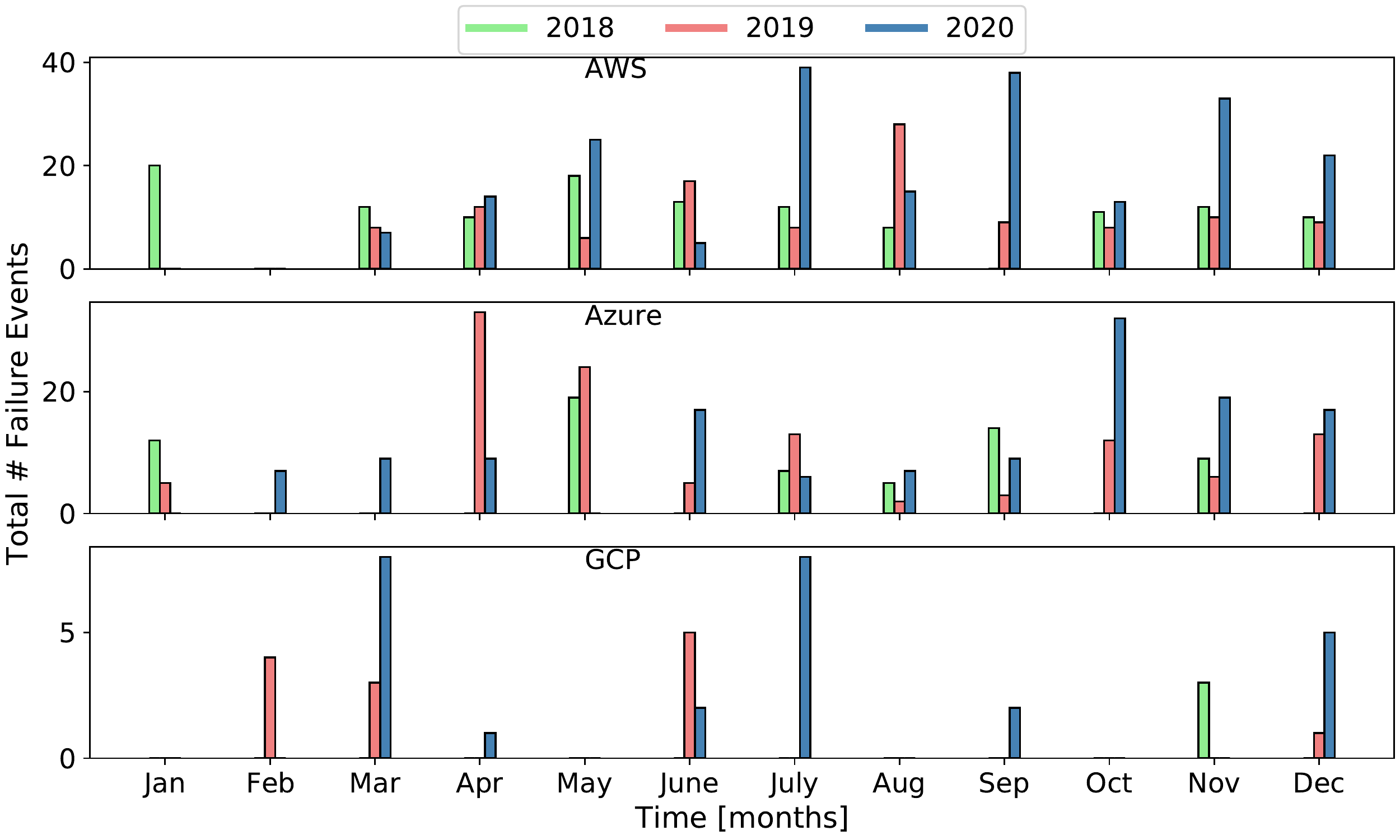}
    \caption{Total number of user-reported cloud failures aggregated per month and year, separated by each cloud vendor}
    \label{fig:user_failure_count}
\end{figure}
The total number of failure events per month and year is depicted in Figure~\ref{fig:vendor_monthly_yearly_failure_count}. The horizontal axis depicts the months per year and the vertical axis depicts the total number of cloud failure events.
GCP experiences fewer failures than AWS and Azure, indicating that either GCP fails less and is more reliable, has fewer users, or users do not use Down Detector to report service failures~(O-\ref{user:failure_event:mf:gcp_less}). 
The hypothesis of GCP fails when provider-reported failures are analyzed. In Table~\ref{tab:provider-detailed-summary} depicts that GCP failed more than Azure and AWS over the past three years regarding provider-reported cloud failures.

There are no failure events in February for AWS. Azure and GCP only experience failure events in that month for 2020 and 2019 respectively. In 2018, only one failure event occurred in GCP.
AWS, Azure, and GCP have respectively, nine, four, and zero months in which they have experienced failure events each year. This changes rapidly for provider-reported failures in Section~\ref{subsec:provider:failure_events}

Moreover, the peaks in Figure~\ref{fig:user_monthly_yearly_average_report_count} and Figure~\ref{fig:user_failure_count} don't match, which implies that the peaks in the number of user reports are caused by certain events that lead lots of users reporting the failure to Down Detector~(O-\ref{user:failure_event:mf:extreme_user_report}). The number of failure events is not proportional to the number of user reports.

\subsection{Failure Duration and Interarrival Times} \label{subsec:user:failure_duration_ia}
In this section, we analyze the distribution of the duration and interarrival times of failure events. Our intention is to gain information on whether failures are short-lived and can be fixed easily or long-lived and might have deep issues. Our goal of investigating the interarrival times is to understand how often cloud services fail and how durable are cloud services in the perspective of users. 
Below are our main findings.

\begin{tcolorbox}
\begin{description}
\itemsep-0.2em
    \mainfinding{analysis:event_trend:mf:gcp_event}{GCP experiences fewer failures with duration times lower and interarrival times higher than AWS and Azure.}
    \mainfinding{analysis:event_trend:high_ia}{Each cloud vendor has periods where it didn't experience a service failure for approximately 100 days.}
\end{description}
\end{tcolorbox}

\begin{figure}[H]
    \centering
    \includegraphics[width=\textwidth]{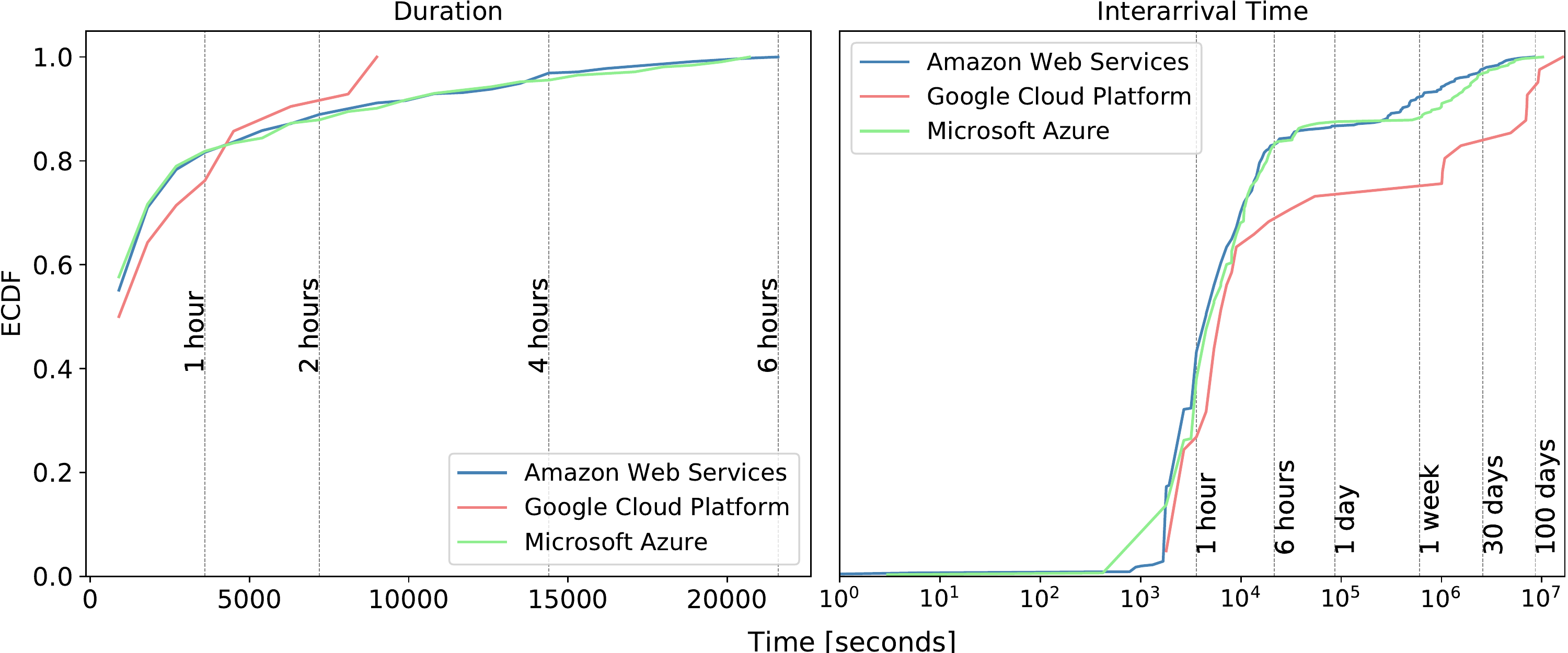}
    \caption{ECDF of failure duration and interarrival time for user-reported cloud failures.}
    \label{fig:duration_ia_ecdf}
\end{figure}
The distribution for duration and interarrival times for failure events is depicted in Figure~\ref{fig:duration_ia_ecdf}. The vertical axis depicts the Empirical Cumulative Distribution Function (ECDF) and the horizontal axis depicts the time in seconds. For each cloud vendor, every point on the plot depicts the fraction of failure duration or interarrival time with the corresponding threshold on the horizontal axis aligned vertically.

A significant amount~(80\%) of the failure events in AWS last below an hour. 
71\% of the failure events for AWS last below 30 minutes. 
75\% of the failure events for AWS last below 45 minutes. 
About 90\% of the failures last less than 2 hours and 15 minutes. 98.6\% of the failures for AWS last less than 5 hours. The longest failure for AWS lasts for 6 hours.
Moreover, failure duration for Azure is similar to Azure. 
75\% of the failure events for Azure are below 45 minutes.  
90\% of the failure events for Azure last less than 2 hours and 30 minutes. The longest failure for Azure lasts for 5 hours and 45 minutes.

Unlike AWS and Azure, GCP experiences shorter failures.
Approximately 75\% of the failure events for GCP are below an hour, and about 90\% of the failures last less than 1 hour and 45 minutes. 
The longest failures (3 failures) last for 2 hours and 30 minutes. 
GCP experiences fewer failures with duration time lower and interarrival time higher than AWS and Azure~(O-\ref{analysis:event_trend:mf:gcp_event}). This indicates that the failures of GCP, in the eyes of the user, have a quick and stable fix, compared to AWS and Azure. AWS and Azure have quite similar duration and interarrival times, nonetheless, some failure events of AWS are slightly longer than Azure. 
The longest failure among all the failures was experienced by AWS with a failure duration of 6 hours. 

The interarrival times for Azure and AWS are similar to each other, yet shorter than GCP. This implies that Azure and AWS fail more often than GCP according to users. In Figure~\ref{fig:duration_ia_ecdf}, up until approximately a week, the interarrival times AWS and Azure overlap with each other. After a week, the ECDF for AWS surpasses Azure.  Approximately 75\% of interarrival times for AWS and Azure are below 3.5 hours.
90\% of the interarrival times for AWS and Azure are below 5.5 days and 11.5 days, respectively. 

The longest interarrival times for AWS, Azure, and GCP are 96.4, 118.5, and 187.9 days~(O-\ref{analysis:event_trend:high_ia}), respectively. Thus, compared to Azure, AWS fails more often and with slightly higher duration. Azure and AWS both fail more often and longer than GCP. 

\section{Analysis of Provider-Reported Cloud Failures}
\label{sec:provider_analysis}
In this section, we analyze the data that was scraped from cloud vendor status pages~\cite{aws_status_page, gcp_status_page, azure_status_page} of GCP, AWS, and Azure to address RQ2 and RQ3.
Similar to the user-reported analysis, the provider-reported analysis gives information on failure event count, duration, and interarrival time. We also investigate the location and type of service failures of provider-reported cloud failures. 
While investigating provider-reported cloud failures, we compare user- and provider-reported cloud failures in terms of failure event count, duration, and interarrival time. 
Our main goal is to compare user-reported failures with the ground truth, provider-reported failures, and understand how failure events are reported by users and what is the difference between the identity of a cloud failure between user- and provider-reported failures.
In this section, there are seven main findings in total.
Main findings are presented at the beginning of each section.

\begin{table}[ht]
\centering

\setlength{\fboxsep}{0pt}
\resizebox{\textwidth}{!}{
\colorbox{lightgray}{
\begin{tabular}{lrr}
\toprule
 Cloud service &  \# failure events & Date range \\
\midrule
     Google Cloud Platform  &  442 & 3 Jan. 2018 - 14 Dec. 2020 \\
     Microsoft Azure &       211 & 14 Jan. 2018 - 21 May 2020\\
     Amazon Web Services &   380 &  4 Jan. 2018 - 18 Dec. 2020\\

\midrule
    Total & 1003 \\
\bottomrule
\end{tabular}}
}
\caption{\label{tab:provider-detailed-summary}Detailed summary for Provider-Reported Cloud Failures, separated by cloud services.}
\end{table}

\subsection{Failure Events}
\label{subsec:provider:failure_events}
In this section, we observe the number of provider-reported cloud failure events per month and year. 
Total number of failure events is summarized in Table~\ref{tab:provider-detailed-summary}.
We compare user- and provider-reported failure event counts and aim to find out whether all failure events are reported by users and are there significant differences between user- and provider-reported failure event counts. 
We also investigate whether provider-reported failure events overlap with user-reported failure events for GCP.  
Below are our main observations.

\begin{tcolorbox}
\begin{description}
\itemsep-0.2em
    \mainfinding{provider:event_count:mf:uniform}{Compared to user-reported failure events, provider-reported failure events are more uniformly distributed throughout the months and years.}
    \mainfinding{provider:event_count:mf:gcp_flying}{GCP has the highest failure event count in provider-reported failure events, whereas GCP has the lowest failure event count in user-reported failure events.}
    \mainfinding{provider:event_count:mf:overlap}{Twelve provider- and user-reported failure events for GCP overlap by event start time.}
    \mainfinding{provider:event_count:mf:less_failure}{Cloud services fail less on the weekend}
\end{description}
\end{tcolorbox}

\begin{figure}[H]
    \centering
    \includegraphics[width=\textwidth]{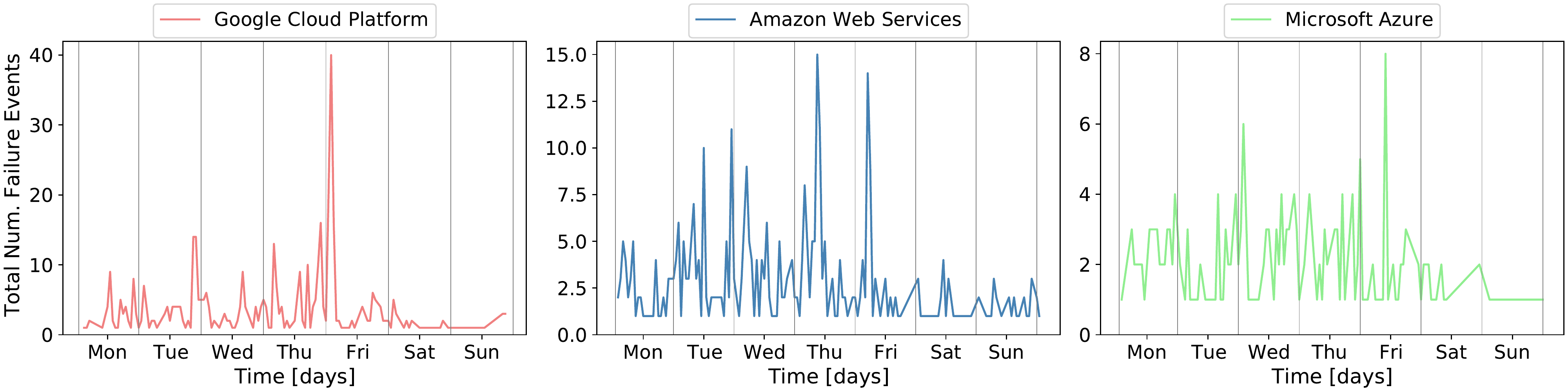}
    \caption{Distribution of provider-reported failure events by hour of week, separated by cloud vendors.}
    \label{fig:vendor_weekly_total_failure_event}
\end{figure}

The number of failure events aggregated by hour of week is depicted in Figure~\ref{fig:vendor_weekly_total_failure_event}.
The horizontal axis depicts the hours in a week and the vertical axis depicts the total number of cloud failure events per hour of week.
In Figure~\ref{fig:vendor_weekly_total_failure_event}, the number of failure events are less on the weekend for the three cloud vendors.
This matches the weekly user failure reports in Figure~\ref{fig:weekly_total_report_count}, as AWS and Azure have less user reports during the weekend.
The reason for less service failures on the weekend might be that the three cloud vendors offer more business and work related cloud services, and the consumers of the services are not working on weekends. Therefore, the cloud services are being utilized less, leading to less workload and service failures.

\begin{figure}[H]
    \centering
    \includegraphics[width=\textwidth]{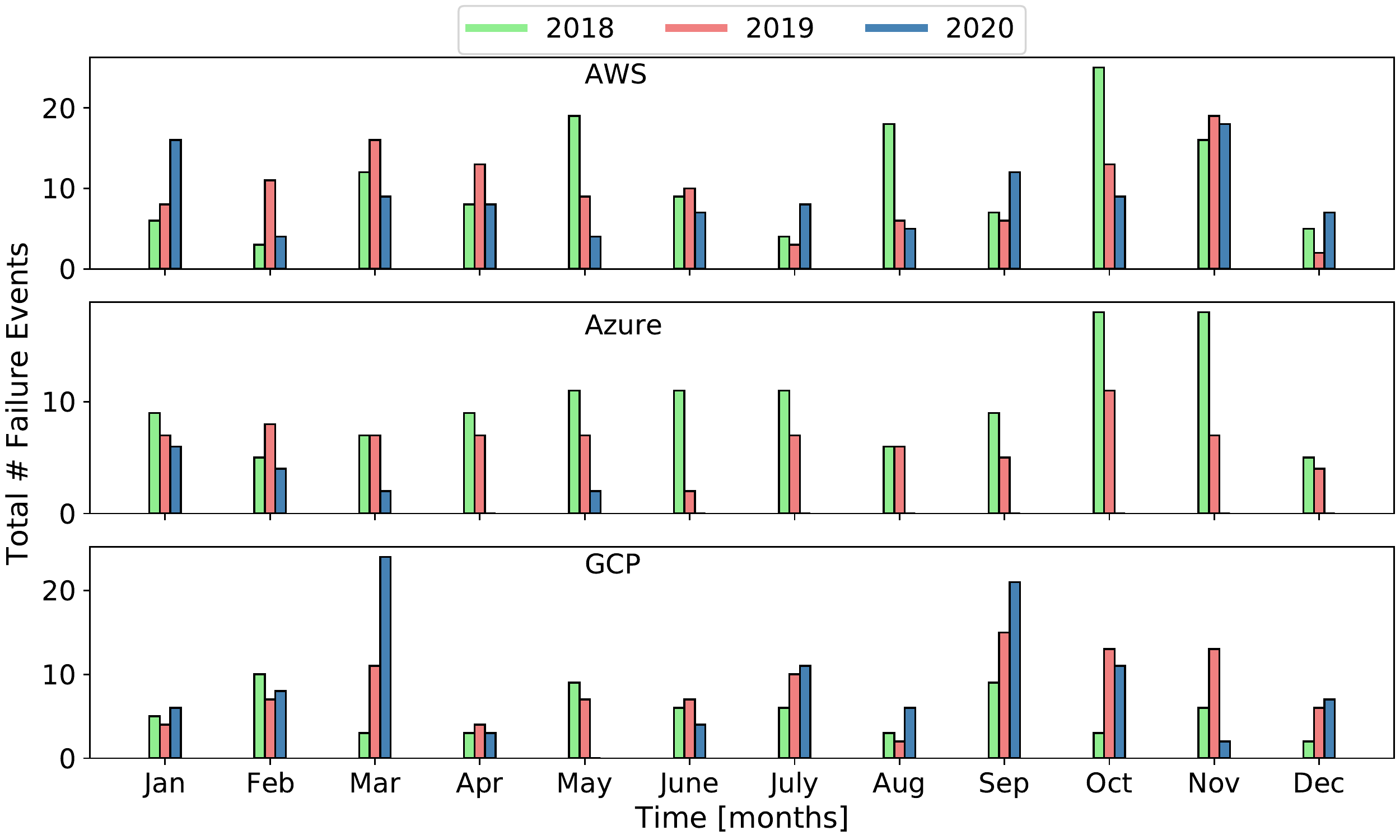}
    \caption{Total number of provider-reported cloud failures per month and year, separated by each cloud vendor.}
    \label{fig:vendor_monthly_yearly_failure_count}
\end{figure}

The total number of failure events per month and year is depicted in Figure~\ref{fig:vendor_monthly_yearly_failure_count}. The horizontal axis depicts the months per year and the vertical axis depicts the total number of cloud failure events.

According to Figure~\ref{fig:vendor_monthly_yearly_failure_count}, the minimum number of failure events during 2020 is experienced by Azure. At most ten failure events have occurred per month and there are no failure events for eight months during 2020 for Azure. 
The maximum number of failure events within GCP, AWS, and Azure is experienced in March 2020, October 2018, and November 2018, respectively.
Unlike user-reported failure events (see Figure~\ref{fig:user_failure_count}), provider-reported failure events are distributed more uniformly throughout the months~(O-\ref{provider:event_count:mf:uniform}). According to user-reported cloud failures, there are no user failure reports of Azure for six months in 2018, GCP for four months across the whole period, and AWS for February for the whole period. 
Contrarily, in provider-reported cloud failures, there is no month in which a failure event does not occur across all three years. 
Moreover, in user-reported cloud failures, GCP experiences the least amount of failure events whereas in provider-reported failures, GCP experiences the most amount of failures~(O-\ref{provider:event_count:mf:gcp_flying}). 
Perhaps not all failure events cause problems in the user experience, or users did not report certain failure events.

\begin{figure}[H]
    \centering
    \includegraphics[width=\textwidth]{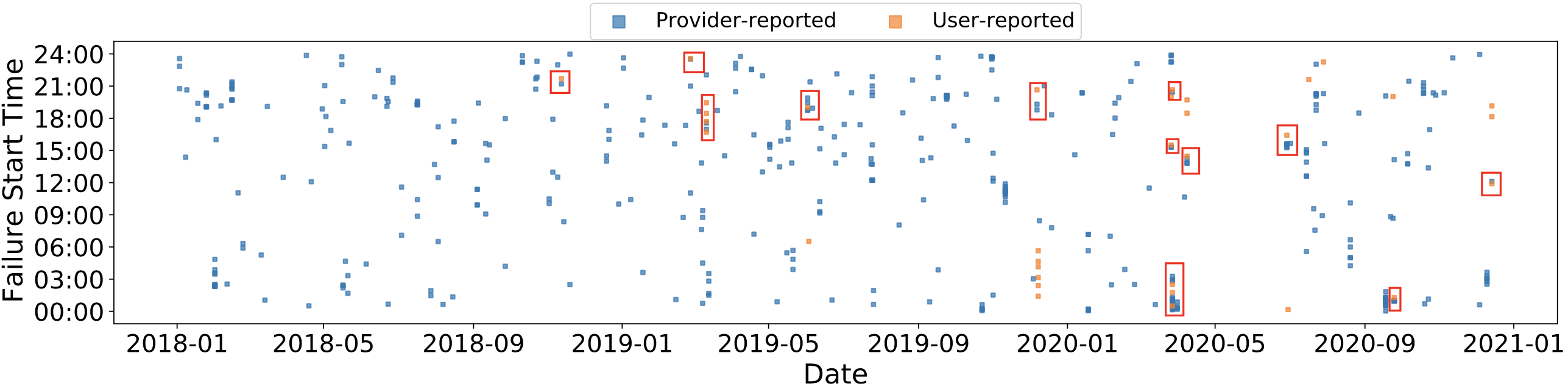}
    \caption{Failures reported by GCP on their official status page compared to failures
reported by users to Down Detector. Overlapping failures are annotated in red
rectangles.}
    \label{fig:gcp_failure_comparison}
\end{figure}

Certain user-reported failure events overlap with provider-reported failure events for all cloud vendors. For simplicity, we only showcase the comparison of user- and provider-reported failures for GCP in Figure~\ref{fig:gcp_failure_comparison}. 
The horizontal axis depicts the date and the vertical axis depicts the time when the failure event started.
According to the figure, some provider- and user-reported failure events for GCP overlap with each other by event start time~(O-\ref{provider:event_count:mf:overlap}). Most of the overlapping failures occur in 2020 and the remaining occur in 2019. There is one overlapping failure event in 2018, which can be expected given that there is only one month in which a failure event occurs in GCP. 
Some user-reported failure events occur right after the provider-reported failure events. The reason might be that a service failure in GCP might not have effected the end-users immediately, and users have reported the failure after they experience failure symptoms.
The graphs for the comparison of AWS and Azure are in the software we used to conduct the analysis~\cite{characerization_scripts}. 

\subsection{Failure Duration and Interarrival Times}
\label{subsec:provider:failure_duration_ia}
We analyze the failure duration and interarrival time to understand how long cloud failures last and how frequently cloud services fail. We intend to compare the duration and interarrival times of user- and provider-reported cloud failures. The differences can reveal information on how much user-reported failure duration and interarrival time match with provider-reported failure duration and interarrival time. 
Our observations are below.

\begin{tcolorbox}
\begin{description}
\itemsep-0.2em
    \mainfinding{provider:mf:last_longer}{Provider-reported failure events last longer than user-reported failures events.}
    \mainfinding{provider:mf:zero_ia}{Each cloud vendor experiences multiple service failures at the same time, leading to instance of zero interarrival times. The data for user-reported cloud failures separates failure events only per vendor.}
\end{description}
\end{tcolorbox}

\begin{figure}[H]
    \centering
    \includegraphics[width=\textwidth]{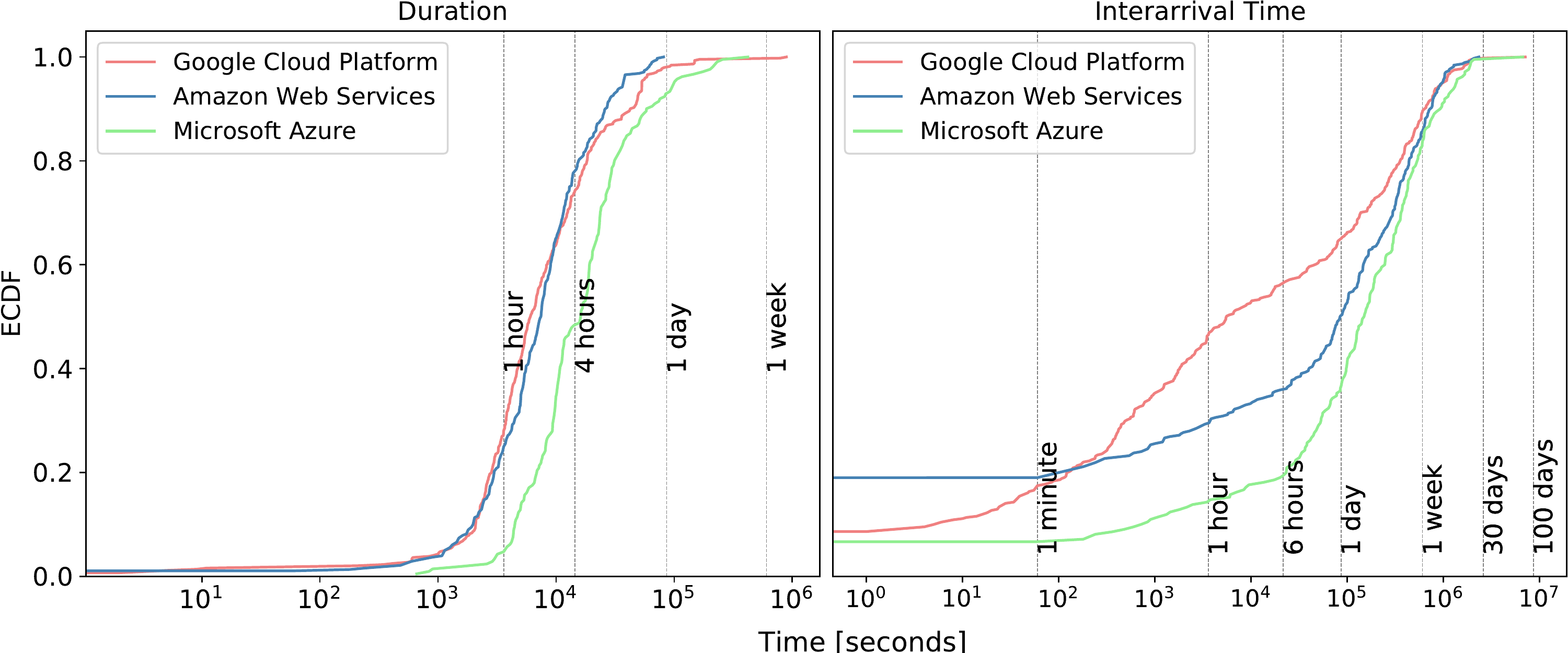}
    \caption{ECDF of failure duration and interarrival time for provider-reported cloud failures.}
    \label{fig:vendor_duration_ia_ecdf}
\end{figure}

Failure duration for user-reported failures is higher than provider-reported failures~(O-\ref{provider:mf:last_longer}). 
Compared to Figure~\ref{fig:duration_ia_ecdf}, there exist failures lasting for more than a week for GCP and Azure in Figure~\ref{fig:vendor_duration_ia_ecdf}. 
The reason can be the monitoring time. Down Detector only monitors a potential failure event for 23 hours and 45 minutes, see Section~\ref{subsec:user-cloud-failures}. Therefore, it is unlikely that failures lasting for more than 23 hours and 45 minutes in Down Detector exist. Also, during a failure that lasts for a week, users might only report a failure event at initial moment of experiencing failure symptoms. Thus, the number user reports might rise at the beginning of the event and quickly decline after the initial phase. 

Each cloud vendor reports their failures per service. If GCP experienced simultaneously two failures in their Networking service and Compute Engine, respectively, they would report two distinct failures with the same failure event start times.
In Figure~\ref{fig:vendor_duration_ia_ecdf}, about 10\% of the failures in GCP and Azure, and 20\% of the failures in AWS have zero interarrival time, indicating that multiple services have failed simultaneously for approximately 10\% of the failures in GCP and Azure, and 20\% of the failures in AWS~\ref{provider:mf:zero_ia}. 
Moreover, the maximum interarrival times are high, approaching 100 days for GCP and Azure.
All cloud vendors, except for AWS, have periods where they did not experience a service failure for approximately 100 days.
The maximum interarrival time for AWS is almost 30 days whereas for user-reported failures, all the cloud vendors exhibit interarrival times up to 100 days. This implies that certain failure events in AWS weren't reported by users, which lead to a 100 day interarrival time.
Nonetheless, in general, the failure duration and interarrival time distribution for user- and provider-reported failures are not similar to each other.

\subsection{Services that Failed} \label{subsec:provider:failed_services}
Different services within cloud vendors can fail. For instance, two services can fail at the same time, leading to simultaneous outages within one vendor. 
In this section, we observe which cloud services in what proportion have failed. 
To get the proportion of each cloud service failure, we aggregate failure events per cloud vendor and divide the proportion of each type of service failure with the aggregated failure events. 
This reveals information on which cloud service failed in what percentage compared to all failures within the cloud vendor.


The highest service failure rate belongs to GCP networking service. Other GCP services like Google Compute Engine, App Engine, Cloud Storage depend on GCP network. Thus, a failure in the networking service can lead to failures in network dependent services. 

GCP offers various services. Each service can fail, either effecting other dependent services or not. Failures in GCP are uniformly distributed among their services. The networking service failed the most  with 11.76\% failure ratio, then comes Google Compute (Infrustructure as a Service) Engine with 8.76\%. Google App Engine is the third most failed service with 7.01\%.  Google Cloud Storage (object storage), Stackdriver (monitoring tool), and Kubernetes Engine(Containers as a Service) all failed in the same rate (6.11\%). 

Service failure types in AWS are also as uniformly distributed as GCP. Majority of the service failures in AWS occur in Amazon Elastic Computing service with 17.89\% failure ratio. Then comes Amazon Relational Database Service, Amazon CloudFront, and Amazon CloudWatch with respectively 5\%, 4.74\%, and 3.95\% failure rates.

In contrast to GCP and AWS, there has been failure events in Azure where all of their services have failed. In fact, majority~(8.06\%) of the cloud failures in Azure effected all their services. Azure is the only vendor which experiences cloud failures where all of their services fail.
Like GCP and AWS, Azure experienced multiple service failures. Similar to AWS, service failures are uniformly distributed among Azure. Moreover, Azure storage failed 5.69\% of the time. Then comes App Service, Azure Portal, and Multiple Services with all having 4.74\% failure ratio. 

\subsection{Failure Locations} \label{subsec:provider:locations}
Cloud vendors have datacenters in various parts of the world. These datacenters host cloud servers and serve the digital society in that location. 
In this section, we want to learn whether datacenters in certain locations experience more failures. 
To get the proportion of each cloud failure location, we aggregate failure events per cloud vendor and divide each location of service failure with the aggregated failure events. This reveals information on what is the percentage of failure location per cloud vendor.
Our observations for this section are below.
\begin{tcolorbox}
\begin{description}
\itemsep-0.2em
    \mainfinding{provider:location:mf:multi_regional}{All cloud vendors experience some of their service failures in multi-regional scale.}
\end{description}
\end{tcolorbox}

In GCP, a significant amount~(35.05\%) of the failures are multi-regional, meaning that multiple regions were effected by the failure. Rest of the failures occurred in the central and east side of the United States and west side of Europe. Minority of the failures occurred in southeast Asia and northern Europe. 

Majority~(35.26\%) of the failures in AWS occurred in Northern Virginia, which seems normal given that the datacenters with the largest capacity are in Virginia, and AWS is planning to build more datacenters~\cite{aws_virginia}. 
Compared to GCP, the amount of multi-regional failures~(10.79\%) are lower in AWS and a small proportion~(5.79\%) of the failures in AWS happened globally. Almost half of the cloud failures~(45.02\%) for Azure occur at multi-regional scale. Similar to AWS, a small amount~(5.21\%) of the failures were at the global scale for Azure. 
All cloud vendors experienced service failures at multi-regional scale~(O-\ref{provider:location:mf:multi_regional}).

\section{Threats to Validity} \label{sec:validity}
In this section, we discuss the threats that could negate the validity of this paper. The main contributors of the user-reported failure data are the users that experience issues in the services they use. The sources for provider-reported failure data are the cloud vendors. Thus, there are some challenges and threats to the validity of this work.

The failure symptoms users are experiencing might be client-side issues, rather than server-side problems. Therefore, some user reports might not be actual cloud service failures. 
Some failure events from the provider- and user-reported failures do not match. Either there is a missing provider-reported failure, indicating that perhaps the cloud vendor did not report a service failure, or a missing user-reported failure, meaning that users did not report a service failure, or they did not experience any failure symptoms and again did not report at all.
Moreover, we manually observed the symptoms for user-reported cloud failures and concluded that they were invalid. Every failure symptom for each cloud vendor had the same symptom type and percentage of experience by users. It is highly unlikely that for the past three years users experience the same symptoms for each service failure.

An issue with Down Detector is that a user can report a failure event twice, which means that some user reports might be doubled by the same user. Furthermore, the number of user reports can decrease rapidly because all the users have reported the service failure at the time they started experiencing failure symptoms and are waiting for the service to work normally. The service, however, is still not functioning properly. 
Therefore, the duration of failure events monitored by Down Detector might not be reliable. Another threat is that the dynamic threshold we used for failure extraction might not be valid for long failure events. Down Detector monitors a failure event for 23 hours and 45 minutes, and there is a short moment where the number of user reports rise and fall, which cannot exceed the monitoring time. Nonetheless, certain failure events prolong a week according to provider-reported failures. 

There might be problems with provider-reported data extraction, especially for Azure. For instance, the Azure status page~\cite{azure_status_page} and the analysis in Section~\ref{sec:provider_analysis} don't match in some cases.


\section{Related Work} \label{sec:related}
In this section, we study previous work that is somewhat similar to ours. 

The work that is most similar to ours is conducted by Gunawi et al.~\cite{gunawi2016does}. They analyze data from public news reports on cloud failures. The data spans seven years and contains outage duration, root causes, impact, and fixing procedures. Their work examines 32 popular cloud services including chat (e.g., WhatsApp), e-commerce (e.g., Ebay), email (e.g., Hotmail), games (e.g., Xbox live), Platform as a Service and Infrastructure as a Service (e.g., Azure), Software as a Service (e.g., Google docs), social (e.g., Twitter), storage (e.g., Dropbox), and video services. (e.g., Netflix). Moreover, their work answers the questions of how long and often outages occur across wide range of Internet services. The majority of the analysis focuses on the root causes of outages. Their analysis of root causes shows that to ensure that cloud services do not have a single point of failure~(SPOF), perfection is required along the whole failure recovery chain: complete failure/anomaly detection, flawless failover code, and working backup components. Yet, many of the outages studied are rooted in some flaws within the failure recovery chain.

Another research, moderately similar to our work, uses provider and user data.
Birke et al.~\cite{birke2014failure} analyzes failures on physical and virtual machines using data reported by users or captured by monitoring tools. Some conclusions are that VMs have lower failure rates and lower probability of failure recurrences than PMs. CPU units, memory size, and memory utilization are the most influential factors for PM failures, whereas CPU counts, the number of disks, and the CPU utilization are the key factors for a VM failure.

Some customer-facing services such as Amazon S3 or MySQL generally use user-reported data to characterize failure reports or other issues that cause the services to not work properly.
Frattini et al.~\cite{frattini2013analysis} analyzes publicly available software bugs of Apache Virtual Computing Lab~(Apache VCL)~\cite{apache_vcl}. The classification and analysis of the bugs are performed based on components, phases, defect types, report time, and relations among them. A contribution of their work is that the approach to analyze the bugs in Apache VCL can be used on other open source cloud platforms. 
Fonseca et al.~\cite{fonseca2010study} studies the internal and external effects of concurrency bugs by examining the user-reported bugs that occurred in MySQL~\cite{mysql}. Their work focuses on the effects of the bugs rather than on their causes.
Yin et al.~\cite{yin2011empirical} studies and analyzes 546 misconfiguration cases in real-world misconfigurations in both commercial and open-source systems. They focus on user-reported software misconfigurations because there is insufficient data for hardware misconfigurations on systems running open-source software.
Fiondella et al.~\cite{fiondella2013cloud} analyzes the cloud incident data reported by companies and news outlets. They provide understanding for different types of failures and their causes and impacts on cloud services. 
Yuan et al.~\cite{yuan2014simple} studies user-reported failures in five popular distributed data-analytic
and storage systems. The goal of their study is to identify failure event sequences to improve the
availability and resilience of the data-analytic
and storage systems. They found that most catastrophic failures are caused by incorrect error handling.
Palankar et al.~\cite{palankar2008amazon} conducts the first independent characterization of Amazon S3 and observes the availability and data access performance using data on user-observed performance. They identified that Amazon S3 wasn't designed for the science community because the science community has specific requirements and challenges regarding data usage that S3 can't address.
Benson et al.~\cite{benson2010first} studies problems experienced by users in an Infrustructure as a Service~(IaaS) platform. They study user problems logged by the open support forum of a large IaaS cloud provider. They examine and classify message threads appearing in the forum over a three-year period, and develop a scientific categorization of problem classes.

Traces provided by Google were used to characterize how various failures occur. 
Rosa et al.~\cite{rosa2015understanding} explores unsuccessful job and task executions using traces of a Google datacenter~\cite{google_cluster_data}. Specifically, they study three types of unsuccessful executions, namely, fail, kill, and eviction. The goal of their study is to provide better understanding of the impact of unsuccessful executions on performance.
Garraghan et al.~\cite{garraghan2014empirical} conducts a statistical analysis of a large-scale heterogeneous production cloud environment using Google cloud traces. They analyze the distribution of failures and the repair times for tasks and servers. Their research understands and quantifies the statistical parameters of cloud failures and repair characteristics to study system behavior as well as providing simulation parameters of cloud computing environments.
Chen et al.~\cite{chen2014failure} studies and characterizes failures from the Google cloud cluster workload traces spanning one month. They work on failure prediction and anomaly detection in cloud applications and aim to improve the dependability of cloud infrastructures by understanding the characteristics of job failures. 

Compared to the works described above, our work contributes an analysis of three big cloud services using data from users gathered by Down Detector and data from cloud vendor status pages.

\section{Conclusion} \label{sec:conclusion}
Cloud computing is the backbone of the digital society. We depend on services offered by cloud systems and the demand is increasing. 
Inevitably, the cloud systems and the services they offer fail, leading to unhappy users and losses in revenue for cloud service providers. To prevent future failures, we intend to understand cloud failures using the failure data from crowdsourcing failure aggregator and cloud vendor status pages.

In this work, we studied how cloud services fail in the perspective of the users, and conducted a study of user- and provider-reported failures spanning three years.
We collected, characterized, and open-sourced failure data from a crowd-sourced failure aggregator. To compare user-reported failures with a reliable source, we characterized provider-reported failure reports and compared the results with the characterization of user-reported failures.
In this work we have: 
(1) identified the challenges associated with gathering and analyzing data from cloud failures, and addressed them through a method focusing on user-reported data; 
(2) characterized patterns in how and when cloud services fail for both user- and provider-reported cloud failures and compared the results with each other; 
(3) and open-sourced a unique long-term failure data from two crowd-sourced failure aggregators, Down Detector and Outage Report, and associated analysis-code.
We have summarized our findings in 13 main observations.
The software and data are available online~\cite{characerization_scripts}.

Future work can compare the characterizations from user-reported failures collected Outage Report and provider-reported failures collected from the status pages of the vendors with each other.
Characterizations of failure reports from Down Detector and Outage Report can be compared with each other.
Failures of client-facing cloud services like Zoom, Netflix, and YouTube can be characterized during and before the COVID-19 period. 


\bibliographystyle{abbrv}
\bibliography{main}

\end{document}